\ifx\pdfoutput\undefined        % not running pdflatex
  \documentclass{kluwer}
  \usepackage{url}
\else                           % running pdflatex
  \documentclass[pdflatex]{kluwer}
\fi

\usepackage{graphicx}
\usepackage{times}
\graphicspath{{./png/}}
\newcommand{\EQ}{\begin{equation}}
\newcommand{\EN}{\end{equation}}
\newcommand{\EQA}{\begin{eqnarray}}
\newcommand{\ENA}{\end{eqnarray}}
\newcommand{\eq}[1]{(\ref{#1})}

\newcommand{\Eq}[1]{Eq.~(\ref{#1})}
\newcommand{\Eqs}[2]{Eqs.~(\ref{#1}) and~(\ref{#2})}
\newcommand{\EEqs}[2]{Equations~(\ref{#1}) and~(\ref{#2})}
\newcommand{\eqs}[2]{(\ref{#1}) and~(\ref{#2})}
\newcommand{\Eqss}[2]{Eqs.~(\ref{#1})--(\ref{#2})}

\newcommand{\Fig}[1]{Figure~\ref{#1}}

\newcommand{\dd}{{\rm d} {}}

\newcommand{\yprl}[5]{: #1, #5, {\em Phys.\ Rev.\ Lett. }{\bf #2}, #3--#4.}

\newcommand{\yphl}[5]{: #1, #5, {\em Phys.\ Lett. }{\bf #2}, #3--#4.}
\newcommand{\yoleb}[5]{: #1, #5, {\em Orig.\ Life Evol.\ Biosph. }{\bf #2}, #3--#4.}
\newcommand{\yjour}[6]{: #1, #6, {\em #2} {\bf #3}, #4--#5.}

\newcommand{\ybook}[3]{: #1, {\em #2}, #3.}

\newcommand{\pjourS}[4]{: #1, #3, {\em #2} (in press,
preprints available online at \url{#4}).}

\newcommand{\poleb}[3]{: #1, #2, {\em Orig.\ Life Evol.\ Biosph.} (in press,
preprints available online at \url{#3}).}

\newcommand{\eaa}{{\em et al.\ }}
\def\half{{\textstyle{1\over2}}}

\newcommand{\const}{\,{\rm const}}

\begin{document}
\begin{article}
\begin{opening}

\title{Dissociation in a polymerization model of homochirality}
\author{A. \surname{Brandenburg}$^*$}
\author{A.~C. \surname{Andersen}}
\author{M. \surname{Nilsson}}
\institute{Nordita, Blegdamsvej 17, DK-2100 Copenhagen \O, Denmark\\
($*$ author for correspondence, e-mail: brandenb@nordita.dk,
phone +45 353 25228, fax: +45 353 89157)}
\runningtitle{Dissociation in a polymerization model of homochirality}
\runningauthor{Brandenburg et al.}
\date{\today,~ $ $Revision: 1.38 $ $}

\begin{abstract}
A fully self-contained model of homochirality is presented that
contains the effects of both polymerization and dissociation.
The dissociation fragments are assumed to replenish the substrate
from which new monomers can grow and undergo new polymerization.
The mean length of isotactic polymers is found to grow slowly with
the normalized total number of corresponding building blocks.
Alternatively, if one assumes that the dissociation fragments themselves
can polymerize further, then this corresponds to a strong source of
short polymers, and an unrealistically short average length of only 3.
By contrast, without dissociation, isotactic polymers becomes infinitely long.
\end{abstract}

\keywords{DNA polymerization, enantiomeric cross-inhibition,
origin of homochirality. $ $Revision: 1.38 $ $}

\end{opening}

\section{Introduction}

Central to the question of the origin of life is the polymerization of
the first complex molecules that can have catalytic properties and that
would eventually carry genetic information.
It is widely accepted that our current life form involving DNA carrying
the genetic code and RNA producing the proteins that, in turn, catalyze
the production of nucleotides, must have been preceded by a simpler
life form called the RNA world
(Woese, 1967; Crick, 1968; Orgel, 1968; see also Wattis \& Coveney 1999).
Here, the RNA has multiple functionality, it carries genetic code and
it is also able to catalyze the production of new nucleotides.

The RNA of all terrestrial life forms involves a backbone of
dextrorotatory (right-handed) ribose sugars.
Theoretically, life could have been equally well based on levorotatory
(left-handed) sugars.
Unless this selection was somehow externally imposed, e.g.\ via circularly
polarized light (Bailey, 2001), magnetic fields (Thiemann 1984),
or via effects involving the parity-breaking
electroweak force (e.g., Hegstrom, 1984), this must have been the result
of some bifurcation process.
Indeed, the homochirality of left-handed amino acids and of right-handed
sugars in living cells can be explained as the result of two combined
effects, auto-catalytic production of similar nucleotides 
during their first polymerization events and a competition
between left- and right-handed nucleotides.
The general idea goes back to early work of Frank (1953), and has been
developed further by Kondepudi and Nelson (1984),
Goldanskii and Kuzmin (1989), Avetisov and Goldanskii (1993) and more
recently by Saito and Hyuga (2004a).
Of particular interest here is the recently proposed detailed
polymerization model of Sandars (2003); see also Brandenburg \eaa
(2005, hereafter referred to as BAHN) and Wattis and Coveney (2005).
The main point of Sandars' model is the assumption that the polymerization
of monomers of opposite handedness terminates further growth on the
corresponding end of the polymer.
This is referred to as enantiomeric cross-inhibition.
Such inhibition makes it generally quite hard for any polymer to grow
successfully.
However, once a polymer has become successful in reaching an appreciable
length, it will have catalytic properties promoting the production of
monomers of the same chirality as that of the catalyzing polymer.

All the polymerization models presented so far ignore the possibility
of polymers breaking at an arbitrary location.
Without this process polymers would, in the homochiral case,
grow to infinite length which is clearly unrealistic.
We begin by discussing a model for the dissociation of isotactic
polymers, where all the building blocks have the same chirality.
Next, we consider the dissociation of polymers whose one end
has already been spoiled with a monomer of the opposite chirality.
We then incorporate the dissociation model into the full polymerization
model of Sandars (2003) and discuss an important modification that is
necessary to prevent the average polymer length from being too short.

\section{Outline of the model}

The model that we are proposing has arisen through the realization that
the obvious generalization of the polymerization model of Sandars (2003),
to include dissociation, leads to two important difficulties.
It was our desire to resolve these problems in a way that seemed
most natural to us, and that involves the least amount of assumptions
and new parameters.
What we came up with is a closed model that is fully self-contained.
As in the original model of Sandars, new monomers of either chirality
are being produced from an achiral substrate.
However, unlike the original model, no external source of the substrate
is required.
Instead, the substrate can be replenished by the ``waste'' generated
by fragmented polymers.

Before we can discuss the dissociation model, let us explain in a
few words the polymerization model of Sandars.
Here, polymers can grow by the addition of monomers that can have either
the same or the opposite chirality, and the corresponding reaction
coefficients are $k_S$ and $k_I$, respectively.
The subscript $S$ indicates that the chirality of both reaction partners
is the {\it same}.
The addition of a monomer of opposite chirality leads to the {\it inhibition}
of further growth at that end of the polymer,
which is indicated by the subscript $I$.
The process of such an inhibition, also referred to as ``enantiomeric
cross-inhibition'', is the single most important aspect of the model
without which there would be no bifurcation from a racemic (i.e.\
equally many right and left handed building blocks) to a homochiral state.

The fragmentation involves a new parameter: the decay rate $\gamma_S$,
at which a polymer can break up anywhere in the chain.
Again, the subscript $S$ refers to the situation where the partners
involved in the bond have the same chirality.
If the chirality is different, we call the decay rate $\gamma_I$, in
analogy to the corresponding reaction coefficient $k_I$ in the original
polymerization model of Sandars (2003).

The perhaps most obvious assumption for dissociation would be to
let the fragments continue to polymerize with new monomers.
This leads to two undesired features of the model.
In the previous case with only polymerization the homochiral equilibrium
had the property that polymers of different lengths are all equally
abundant.
This goes on all the way to infinity.
If we now allow these polymers to break, there is potentially a
catastrophe in that arbitrarily many short polymers can form.
This is also supported by the numerical simulations discussed below.
Furthermore, the numerical solutions show that, even
in the best possible case, the average polymer length never exceeds 3,
which is clearly unrealistically short.
We propose two alternative ways to allow for the formation of longer
chains.
One possibility is to include an additional degradation of polymers
leading to a loss term in the polymerization equations and a corresponding
source term for the substrate.
Another possibility is to recycle the dissociation products into the
substrate without invoking an additional degradation of polymers.
In both cases the total number of building blocks
in the system is constant, so the substrate plays now an integral part of the model.
As a mechanical analogue, we can think of the mass of the substrate as
being similar to potential energy, and the mass of all polymers as being
similar to kinetic energy, such that the total number (corresponding
to the total energy) is conserved.
Thus, not only goes the production of new left and right handed building
blocks at the expense of the substrate, but now the substrate is
being replenished by the dissociation fragments such that the total number
of building blocks (regardless of their chirality) remains constant.

In the following we develop the model step by step.
We first need to show that the mean polymer length is never more than
3 if the fragments are reused for further polymerization.

\section{Developing the dissociation procedure}

Following the basic idea behind Sandars' 
polymerization model we assume the presence
of left and right handed polymers of length $n$, denoted by $L_n$ and
$R_n$, respectively.
We also assume the presence of polymers whose one end has been
spoiled by a reaction with a monomer of opposite chirality.
The resulting polymers of this form are denoted
by $L_nR_1$ and $R_nL_1$.

\subsection{Isotactic dissociation}
\label{short_poly}

We begin with the description of a dissociation model by discussing
isotactic polymers of length $n$, which are assumed to break
(dissociation) at a mean rate $\gamma_S$
(assumed independent of $n$), at position $m$, according to
\EQ
L_n\stackrel{\gamma_S~}{\longrightarrow}L_m + L_{n-m}.
\EN
Here, $L$ refers to left-handed building blocks, but a corresponding equation
is also valid for right-handed polymers, denoted by $R$.
In the present case, the fragments $L_m$ and $L_{n-m}$ will be reused
for further polymerization.
As an example, $L_4$ can break up into two $L_2$, or into
one $L_1$ and one $L_3$, but for the latter there are two possibilities to
do this.
Thus, for $n=4$ there are altogether $n-1=3$ different ways of destroying $L_4$.
This then leads to an evolution equation for the concentration of
polymers, $[L_n]$,
\EQ
{\dd\over\dd t}[L_n]=...+2\gamma_S\sum_{m=n+1}^N[L_m]-(n-1)\gamma_S[L_n],
\label{DepolLn}
\EN
where $n\ge2$, and the last term represents the decrease of the
concentration $[L_n]$ due to the $n-1$ different ways of breaking
up the polymer.
The first term represents the corresponding gain
from breaking up polymers with $m=n-1$ or more building blocks.
The evolution equation for $[L_1]$ has only a gain term
from breaking up polymers of length $n\ge2$, so
\EQ
{\dd\over\dd t}[L_1]=...+2\gamma_S\sum_{n=2}^N[L_n].
\label{DepolL1}
\EN
The absence of any negative terms (sinks) on the right hand side implies
that, if there is only dissociation, $[L_1]$ can only grow.
The dots in \Eqs{DepolLn}{DepolL1}
denote the possible presence of extra terms (discussed in the
next subsection) that would be needed
to model the primary polymerization process.

\begin{figure}[t!]\begin{center}
\includegraphics[width=\textwidth]{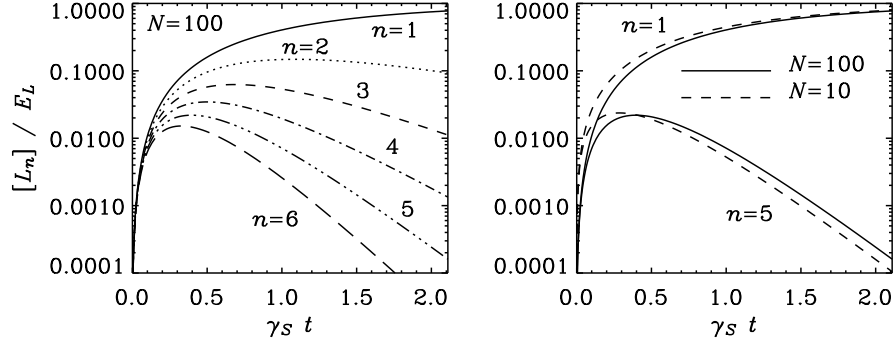}
\end{center}\caption[]{
Evolution of $[L_n]$, using as initial condition
$[L_{100}]=1$ and $[L_n]=0$ for $n\neq0$.
}\label{pbreak10}\end{figure}

The same set of equations \eqs{DepolLn}{DepolL1} applies also to $R_n$.
The mean rate of dissociation is again $\gamma_S$, so the model
is completely symmetric with respect to exchanging $L\rightleftarrows R$.
Using the identity
\EQ
\sum_{n=1}^Nn\sum_{m=n+1}^N[L_n]=\sum_{n=1}^N\half(n-1)n[L_n],
\EN
one can easily see that these reaction equations
\eqs{DepolLn}{DepolL1}, in the absence of extra terms, conserve the total
number of left and right handed building blocks, i.e.\
\EQ
E_L=\sum_{n=1}^N n[L_n]=\const,\quad
E_R=\sum_{n=1}^N n[R_n]=\const.
\EN
As an illustration we show in \Fig{pbreak10} a numerical integration of
the evolution of $[L_n]$, using as initial condition
$[L_{100}]=1$ and $[L_n]=0$ for $n\neq100$.
Thus, we have $E_L=100$ initially, and this value is preserved by the
model for all times.

As can be seen from \Fig{pbreak10}, both monomers and short polymers
are immediately being produced.
For $n\ge2$ the concentration reaches a maximum at a time that is
of the order $\gamma_S^{-1}$, and decays then exponentially to zero.

\subsection{Semi-spoiled polymers}

For polymers whose one end has been spoiled by a monomer of opposite
chirality, we have two types of reactions: those where the
spoiling enantiomer breaks off (rate $\gamma_I$) and those where
the polymer breaks up somewhere else in the isotactic part (rate $\gamma_S$).
Thus, we assume
\EQ
L_nR_1\stackrel{\gamma_I~}{\longrightarrow}L_n + R_1,
\EN
and
\EQ
L_nR_1\stackrel{\gamma_S~}{\longrightarrow}L_m + L_{n-m}R_1
\label{ReactLnR1}
\EN
for $1\leq m\leq n-1$.
Ignoring a particular complication that will be discussed in a moment,
our {\it preliminary} set of equations for these additional reactions
is then given by
\EQ
{\dd\over\dd t}[L_nR_1]=...+\gamma_S\sum_{m=n+1}^N[L_mR_1]
-\left\{\gamma_I+(n-1)\gamma_S\right\}[L_nR_1],
\label{PrelimLnR1}
\EN
\EQ
{\dd\over\dd t}[L_n]=...+\gamma_S\sum_{m=n+1}^N[L_mR_1]
+\gamma_I[L_nR_1],
\label{PrelimLn}
\EN
\EQ
{\dd\over\dd t}[L_1]=...+\gamma_S\sum_{n=2}^N[L_nR_1]
+\gamma_I[L_1R_1],
\label{PrelimL1}
\EN
\EQ
{\dd\over\dd t}[R_1]=...+\gamma_I\sum_{n=1}^N[L_nR_1].
\EN
These equations ignore the dissociation of isotactic polymers
discussed in the previous section, but they can simply be added to the
present set of equations.
Again, the system of equations has to be completely symmetric with
respect to exchanging $L\rightleftarrows R$.
However, the reaction \eq{ReactLnR1} for $m=n-1$ produces $L_1R_1$
at a rate that is proportional to $[L_nR_1]$.
In general, since $[L_nR_1]\neq[R_nL_1]$, this would lead to
$[L_1R_1]\neq[R_1L_1]$, which is not permitted.
We therefore have to discard the reaction \eq{ReactLnR1} for
$m=n-1$, i.e.\ we have to discard the reactions
\EQ
L_nR_1\stackrel{\gamma_S~}{\longrightarrow}L_{n-1}+L_1R_1
\quad\mbox{(discarded)},
\EN
and likewise for the dissociation of $R_nL_1$.
Since we have therefore one reaction less, this means that in
\Eq{PrelimLnR1}, which now applies only for $n\geq2$, the $n-1$ factor
changes effectively into a $n-2$ factor.
Furthermore, in \Eqs{PrelimLn}{PrelimL1} the sums start only with $m=n+2$
and $n=3$, respectively.

When writing down the full set of equations we have to treat the evolution
of $[L_1R_1]$ separately, so
\EQ
{\dd\over\dd t}[L_1R_1]=...-\gamma_I[L_nR_1],
\label{DepolL1R1}
\EN
while for $n\ge2$ we have a pair of equations
\EQ
{\dd\over\dd t}[L_nR_1]=...+w_n^{(LR)}
-\left\{\gamma_I+(n-2)\gamma_S\right\}[L_nR_1],
\EN
\EQ
{\dd\over\dd t}[R_nL_1]=...+w_n^{(RL)}
-\left\{\gamma_I+(n-2)\gamma_S\right\}[R_nL_1].
\EN
Here,
\EQ
w_n^{(LR)}=\gamma_S\sum_{m=n+1}^N[L_mR_1],\quad
w_n^{(RL)}=\gamma_S\sum_{m=n+1}^N[R_mL_1],
\EN
are all the terms that have resulted from dissociation.
The corresponding pair of equations for $[L_n]$ and $[R_n]$ is
automatically valid also for $n=1$, so we have
\EQ
{\dd\over\dd t}[L_n]=...+w_n^{(L)}
-(n-1)\gamma_S[L_n],
\label{DepolLnNEW}
\EN
\EQ
{\dd\over\dd t}[R_n]=...+w_n^{(R)}
-(n-1)\gamma_S[R_n],
\label{DepolRnNEW}
\EN
where
\EQ
w_n^{(L)}=
2\gamma_S\!\!\!\!\sum_{m=n+1}^N\!\![L_m]
+\gamma_S\!\!\!\!\sum_{m=n+2}^N\!\![L_mR_1]
+\gamma_I[L_nR_1]+\delta_{n1}\!\sum_{m=1}^N[R_mL_1],
\label{wL}
\EN
\EQ
w_n^{(R)}=
2\gamma_S\!\!\!\!\!\sum_{m=n+1}^N\!\![R_m]
+\gamma_S\!\!\!\!\!\sum_{m=n+2}^N\!\![R_mL_1]
+\gamma_I[R_nL_1]+\delta_{n1}\!\sum_{m=1}^N[L_mR_1].
\label{wR}
\EN
Here, $\delta_{n1}=1$ for $n=1$, and $\delta_{n1}=0$ for $n\ge2$.

We have calculated solutions using as initial condition $[L_nR]=1$ for
different values of $n$ and found that the evolution of $[L_n]$
is very similar to that shown in \Fig{pbreak10}, so we do not need
to reproduce this result here.

\subsection{Polymerization and dissociation}

\begin{figure}[t!]\begin{center}
\includegraphics[width=\textwidth]{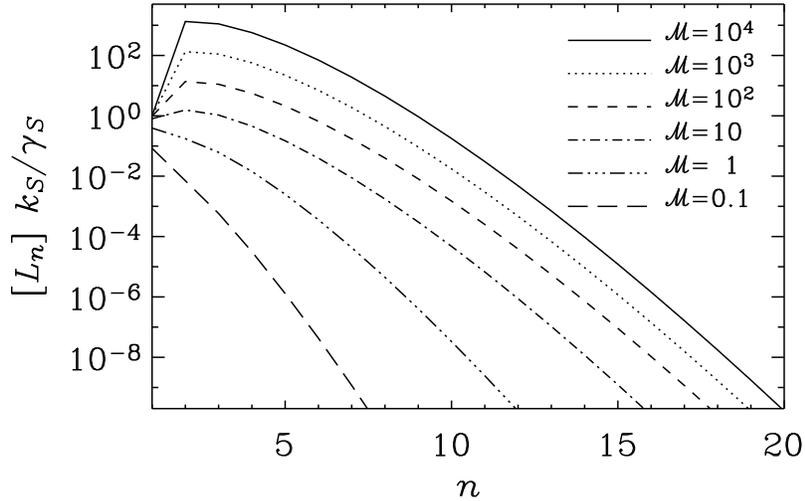}
\end{center}\caption[]{
Isotactic equilibrium states with polymerization and dissociation,
for different values of the universal parameter ${\cal M}$ changing
over a range of six orders of magnitude.
}\label{ppoldepolcomp}\end{figure}

We now add the polymerization equations of Sandars (2003) to
\Eqss{DepolL1R1}{DepolRnNEW}.
Again, we begin by discussing first the homochiral case.
In that case we have only two reactions,
\EQA
L_{n-1}+L_1&\stackrel{2k_S~}{\longrightarrow}&L_{n},\\
L_n&\stackrel{\gamma_S~}{\longrightarrow}&L_m + L_{n-m},
\ENA
where $k_S$ is the reaction coefficient for attaching a monomer
with the same handedness.
The factor 2 on $k_S$ indicates that polymerization can proceed on
both ends of the polymer.
This agrees with earlier approaches, and may be realistic for PNA
polymerization, but not for RNA or DNA polymerization which usually proceeds
only on one end.
Since the monomer can be attached to any one of the two ends of the
polymer, the overall reaction proceeds with the coefficient $2k_S$.
The full reaction equations for the homochiral case can then be written as
(for $n\ge3$)
\EQ
{\dd[L_{n}]\over\dd t}=2k_S[L_1]\Big([L_{n-1}]-[L_{n}]\Big)
+2\gamma_S\!\!\!\sum_{m=n+1}^N\!\![L_m]-(n-1)\gamma_S[L_n],
\label{PolDepolLn}
\EN
while for $n=2$ we have
\EQ
{\dd[L_{2}]\over\dd t}=k_S[L_1]\Big([L_{1}]-2[L_{2}]\Big)
+2\gamma_S\sum_{m=3}^N [L_m]-\gamma_S[L_2],
\label{PolDepolLn2}
\EN
where an extra 1/2 factor has occurred in front of the $[L_1]^2$ term.
(For $n=2$, two pieces of the same species react with each other, whereas
for $n>2$ there are always two different species, i.e.\ monomers and
polymers; see BAHN for a more detailed discussion).
For $n=1$ we have
\EQ
{\dd\over\dd t}[L_1]=-2k_S[L_1]\sum_{n=1}^{N-1}[L_{n}]
+2\gamma_S\sum_{n=2}^N[L_n].
\label{PolDepolL1}
\EN
Note that this problem is governed by three parameters, $k_S$, $\gamma_S$,
and the conserved quantity $E_L$, which only depends on the initial condition.
These three parameters can be combined into a single non-dimensional
parameter,
\EQ
{\cal M}=E_L k_S/\gamma_S
\quad\mbox{(homochiral case)},
\label{calEL}
\EN
that characterizes all possible solutions.
Moreover, these equations possess a unique equilibrium state which
is in general different for different values of ${\cal M}$; see
\Fig{ppoldepolcomp}.
Here we have normalized $[L_n]$ in terms of $\gamma_S/k_S$ to
make it dimensionless.

\begin{figure}[t!]\begin{center}
\includegraphics[width=\textwidth]{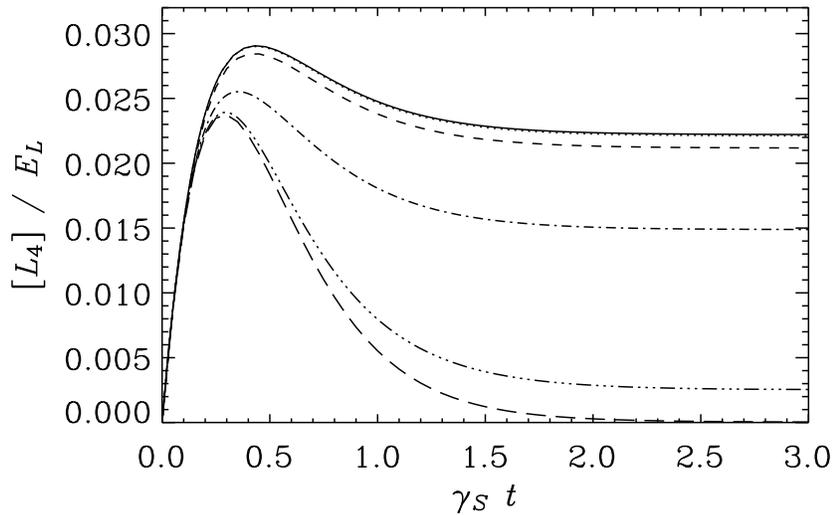}
\end{center}\caption[]{
Relaxation phase toward the isotactic equilibrium states in
the presence of polymerization and dissociation,
for the same six different values of the parameter ${\cal M}$
as in \Fig{ppoldepolcomp}.
}\label{ppoldepolt}\end{figure}

Given that there is a non-dimensional parameter (${\cal M}$)
in the problem, there
is no unique choice for a non-dimensional representation of time.
Possible non-dimensional combinations are $\gamma_S t$ (as used in
\Fig{pbreak10}) and $E_L k_S t$.
In \Fig{ppoldepolt} we show the time dependence of $[L_4]$
(normalized by $E_L$) as a function of $\gamma_S t$ toward the equilibrium
solution shown in \Fig{ppoldepolcomp}.
Note that the approximate position of the maximum is always at around
the same value of $\gamma_S t$ for values of ${\cal M}$ changing
over six orders of magnitude.
This shows that the typical relaxation time scale is governed by
$\gamma_S^{-1}$.

\begin{table}[b!]\caption{   
Mean polymer length $N_L$ for different values of ${\cal M}$,
for isotactic polymers (here left-handed).
}\vspace{12pt}{\begin{tabular}{l|ccccccc}
${\cal M}$ & $0.1$ & $1$ & $10$ & $10^2$ & $10^3$ & $10^4$ \\
\hline
$N_L$        & 1.09 & 1.55 & 2.45 &  2.92  &  2.99  & 3.00 
\label{chainlength}\end{tabular}}\end{table}

It is somewhat surprising that with dissociation, $[L_n]$ always
peaks at small values of $n$ (at $n=2$ or ${\cal M}\ge10$ or at $n=1$
for smaller values of ${\cal M}$).
This can be quantified in terms of the mean polymer length $N_L$ that
can be defined as $N_L=\sum n[L_n]/\sum [L_n]$ (see BAHN).
The resulting values of $N_L$ approach 3 for large values of ${\cal M}$,
but are otherwise always less than 3.

\subsection{Coupling to a substrate}

It is natural to proceed as in the model of Sandars (2003) and couple
the polymerization equation to a substrate from which new monomers can
be produced in a catalytic fashion.
It is sufficient to discuss first the isotactic case, so we add a source $Q_L$
to the right hand side of \Eq{PolDepolL1},
\EQ
{\dd\over\dd t}[L_1]=Q_L-2k_S[L_1]\sum_{n=1}^{N-1}[L_{n}]
+2\gamma_S\sum_{n=2}^N[L_n],
\label{PolDepolL1S}
\EN
where $Q_L$ quantifies the source of new left-handed monomers.
Since this term provides of source of left-handed building blocks,
$E_L$ is no longer conserved.
Instead, as discussed by BAHN, $E_L$ obeys the evolution equation
\EQ
{\dd E_L\over\dd t}=Q_L-2k_S[L_1][L_N].
\EN
In the absence of dissociation, a homochiral steady state is
possible, where $[L_n]$ is constant for all $n\ge2$, so $[L_N]$
is finite and $Q_L$ is balanced by $2k_S[L_1][L_N]$.

Obviously, $Q_L$ should depend on the concentration of the substrate, $[S]$,
so it is natural to write $Q_L=k_C[S]C_L$, where $C_L$ determines the efficiency
of the production of left-handed monomers from the substrate.
Since this generation is supposed to be a catalytic process, $C_L$
should depend in some way on $[L_n]$ itself; here we assume $C_L=E_L$,
but different proposals have been made in the past (see BAHN for a
discussion).
The substrate itself obeys an evolution equation of the form
\EQ
{\dd[S]\over\dd t}=Q-(Q_L+Q_R),
\EN
where $Q$ is a source for the substrate, and $Q_R$ in the present case.
For the moment, this source can be thought of as being externally given,
as in the model of Sandars (2003), but we will assume that this comes
actually from the dissociation fragments.

\begin{figure}[t!]\begin{center}
\includegraphics[width=\textwidth]{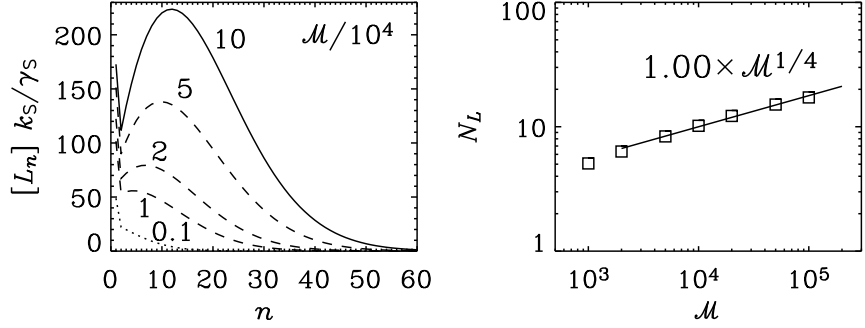}
\end{center}\caption[]{
Isotactic equilibrium states with polymerization, dissociation,
and uniform degradation, for different values of ${\cal M}/10^4$ (left),
and the mean polymer length $N_L$ (right), for $\gamma/\gamma_S=20$.
}\label{pdegrade}\end{figure}

Regardless of the particular choice, we face a general problem in that
dissociation causes the polymers to have finite length, so $[L_N]\to0$
and hence no equilibrium state is possible any more.
This causes a secular (linear) growth, so at some point the numerical
integration develops an arithmetic overflow.
An obvious way to balance this secular growth is to add a simple loss term,
$\dd[L_n]/\dd t=...-\gamma[L_n]$, where $\gamma$ is the degradation rate
and the dots denote all the other terms that are already present.
The result is show in \Fig{pdegrade}.

\subsection{Feeding the fragments back into the substrate}
\label{FragmentsToSubstrate}

Clearly, the dissociation model developed so far requires some
modi\-fications that are necessary to prevent the model from displaying
secular growth when combined with the polymerization model of
Sandars (2003) and to allow for an average polymers length of more than 3.
One possibility would be to make the decay rate $\gamma_S$ dependent on
$n$, for example in such a way that $\gamma_S=0$ for small values of $n$.
One could also think of adding an overall loss term.
Yet another possibility, that is close to our final proposal, is to recycle
the monomers resulting from dissociation back into the substrate.
In the end, however, we found it most plausible to assume that all
fragments resulting from dissociation are recycled back into the
achiral substrate.
Thus, the source term would then be
\EQ
Q=W_L+W_R+W_{LR}+W_{RL}+W_{RLR}+W_{LRL}
\EN
where
\EQ
W_L=\sum_{n=1}^N n w_n^{(L)},\quad
W_R=\sum_{n=1}^N n w_n^{(R)},
\EN
is the total number of recycled building blocks (both left-handed and
right-handed),
\EQ
W_{LR}=\sum_{n=1}^N (n+1) w_n^{(LR)},\quad
W_{RL}=\sum_{n=1}^N (n+1) w_n^{(RL)}
\EN
are the corresponding contributions from fragmented semi-spoiled polymers,
and
\EQ
W_{RLR}=\sum_{n=2}^N (n+2) [R_1][L_nR],\quad
W_{LRL}=\sum_{n=2}^N (n+2) [L_1][R_nL]
\EN
are the contributions from terminally spoiled chains.
The new system of equations is then
\EQ
{\dd\over\dd t}[L_n]=p_n^{(L)}-(n-1)\gamma_S[L_n],
\label{DepolLn_new}
\EN
\EQ
{\dd\over\dd t}[R_n]=p_n^{(R)}-(n-1)\gamma_S[R_n],
\label{DepolRn_new}
\EN
\EQ
{\dd\over\dd t}[L_nR_1]=p_n^{(LR)}
-\left\{\gamma_I+(n-2)\gamma_S\right\}[L_nR_1],
\label{DepolLnR_new}
\EN
\EQ
{\dd\over\dd t}[R_nL_1]=p_n^{(RL)}
-\left\{\gamma_I+(n-2)\gamma_S\right\}[R_nL_1],
\label{DepolRnL_new}
\EN
where $p_n^{(L)}$, $p_n^{(R)}$, $p_n^{(RL)}$, and $p_n^{(LR)}$
indicate the terms due to polymerization
(see Sandars 2003, BAHN, Wattis \& Coveney 2005); see the appendix.
\EEqs{DepolLn_new}{DepolRn_new} are valid for all $n\ge1$, but
\Eqs{DepolLnR_new}{DepolRnL_new} are only valid for $n\ge2$.
For $n=1$ these equations reduce to
\EQ
{\dd\over\dd t}[L_1R_1]=p_1^{(LR)}-\gamma_I[L_1R_1].
\EN
We note that $[R_1L_1]=[L_1R_1]$.
These equations are constructed in such a way that the total
number (or mass) of right and left handed building blocks is
conserved, i.e.\
\EQ
M\equiv[S]+E_R+E_L+E_R^++E_L^+=\const.
\EN
Here, $E_R^\pm=\sum(n\pm1)[R_n]$ and $E_L^\pm=\sum(n\pm1)[L_n]$ have
been introduced.
We recall that due to recycling of the right and left handed building
blocks through an achiral substrate, the total chirality, which involves
$E_R^-$ and $E_L^-$, is not conserved; see Section~5 of BAHN.
The quantity $M$ can be expressed in non-dimensional form,
\EQ
{\cal M}=Mk_S/\gamma_S
\quad\mbox{(general case)},
\EN
which is conserved for all times.
This is therefore the main control parameter of our model.
It generalizes our earlier definition for 
the fully homochiral cases; see \Eq{calEL}.

\begin{figure}[t!]\begin{center}
\includegraphics[width=\textwidth]{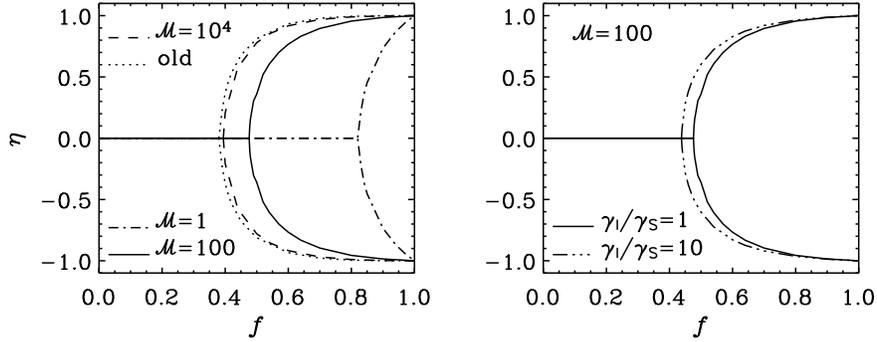}
\end{center}\caption[]{
The effects of ${\cal M}$ and $\gamma_I$ on the bifurcation diagram.
Increasing the values of ${\cal M}$ and $\gamma_I$ allow near-homochiral
states with decreased fidelity.
}\label{bifurc}\end{figure}

In \Fig{bifurc} we show that increasing the value of ${\cal M}$ leads
to an increased range over which the racemic solution is unstable and
a near-homochiral state emerges.
Likewise, increasing the rate at which the spoiling monomers break off
also broadens the range of permissible values of the fidelity for which
the racemic solution is unstable.

\begin{figure}[t!]\begin{center}
\includegraphics[width=\textwidth]{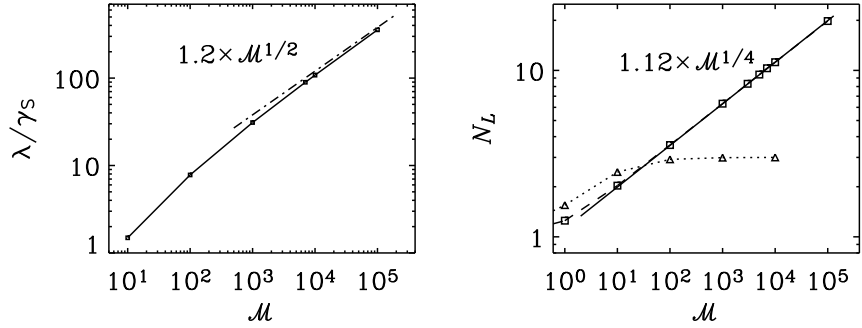}
\end{center}\caption[]{
Dependence of normalized growth rate $\lambda/\gamma_S$ for the racemic
solution and mean polymer length $N_L$ for the homochiral solution
on the total normalized mass parameter ${\cal M}$ for $f=1$,
$k_I/k_S=1$ and $\gamma_I/\gamma_S=1$.
}\label{plam}\end{figure}

When $f$ exceeds a critical value, the enantiomeric excess
\EQ
\eta = \frac{E_{L}-E_{R}}{E_{L} + E_{R}}
\EN
increases exponentially with time like $e^{\lambda t}$, where $\lambda$
is the growth rate.
For $f=1$, the growth rate is (see \Fig{plam})
$\lambda=1.2{\cal M}^{1/2}\gamma_S\equiv(Mk_S\gamma_S)^{1/2}$, so it is
the geometrical mean between the polymerization rate $Mk_S$ on the one
hand and the dissociation rate $\gamma_S$ on the other.

Contrary to the model without recycling, the present model does allow
for polymer lengths that can easily exceed the previous bound of 3.
This requires large values of ${\cal M}$; see \Fig{plam}, where
we plot the resulting values of $N_L$ as a function of ${\cal M}$.
In fact, we find that to a good approximation,
\EQ
N_L\quad\mbox{or}\quad N_R\approx1.12{\cal M}^{1/4}
\quad\mbox{(for ${\cal M}\ge10$)};
\EN
see \Fig{plam}.
We regard the possibility of long chains as a crucial property of any
reasonable polymerization model.
Furthermore, the fact that the model is now fully self-contained
($M$ is conserved) makes it an appealing alternative to previous models.

\section{Conclusions}

Dissociation of polymers appears to be an important component of any
polymerization model.
The present work has shown, however, that the straightforward usage of
dissociation fragments for further polymerization does not yield
realistic model behavior, because the maximum polymer length would
not be more than 3.
Various other modifications that could allow for longer polymers have
been discussed, and it is likely that there are more possibilities.
The main problem is that the fragments from dissociation tend
to produce excessive amounts of short polymers that cause the average
polymer length to be very short.
Consequently, we have postulated that the fragments resulting from
polymerization are instead recycled into the substrate.
The average polymer length then depends on the normalized
dissociation time.
In this model, no external source of the substrate is required, so
the model is now fully self-contained.

The model is governed by the total number of left and right handed
homochiral building blocks, the reaction rates for polymerization with the
same and the opposite chirality, and the corresponding dissociation
rates.
These numbers can be combined into a single non-dimensional number
that characterizes the behavior of the system.
At the moment we have no clear idea about its value,
but laboratory experiments should
be able to determine not only this coefficient, but they should also
allow us to test various aspects and predictions of the model.

We recall that in order to draw conclusions about the time scale on
which homochirality can be achieved, it is important to discuss the
spatial extent of the system (Saito and Hyuga 2004b).
Homochirality may develop rapidly at one point in space, but the
handedness may be different at different locations.
The relevant time scale for achieving global homochirality is
therefore much longer and is given either by the diffusion time scale,
which is very long, or by a turbulent turnover time which can be much
shorter if turbulent flows are present (Brandenburg \& Multam\"aki 2005).
Obviously, the generalizations given in the present paper can directly
be applied to their model provided the local growth remains still
large enough.
%AB: need to quantity...

\appendix
\section{Polymerization terms}

In this appendix we state the terms describing the polymerization process.
These terms are equivalent to those given and discussed in BAHN,
Equations~(20)--(27).
For $n\ge2$ we have
\EQ
p_n^{(L)}=2k_S[L_1]\Big(\sigma_n^{(1/2)}[L_{n-1}]-[L_{n}]\Big)
-2k_I[L_{n}][R_1],
\EN
\EQ
p_n^{(R)}=2k_S[R_1]\Big(\sigma_n^{(1/2)}[R_{n-1}]-[R_{n}]\Big)
-2k_I[R_{n}][L_1],
\EN
\EQ
p_n^{(RL)}=k_S[R_1]\Big(\sigma_n^{(0)}[R_{n-1}L]-[R_{n}L]\Big)
+k_I[L_1]\Big(2[R_{n}]-[R_nL]\Big),
\EN
\EQ
p_n^{(LR)}=k_S[L_1]\Big(\sigma_n^{(0)}[L_{n-1}R]-[L_{n}R]\Big)
+k_I[R_1]\Big(2[L_{n}]-[L_nR]\Big),
\EN
whereas for $n=1$ we have $p_1^{(L)}=-\lambda_L[L_1]$ and
$p_1^{(R)}=-\lambda_R[R_1]$, where
\begin{equation}
\lambda_L=
2k_S\sum_{n=1}^{N-1}[L_n]
+2k_I\sum_{n=1}^{N}[R_n]
+k_S\sum_{n=2}^{N-1}[L_nR]
+k_I\sum_{n=2}^{N}[R_nL],
\label{lambdaL}
\end{equation}
\begin{equation}
\lambda_R=
2k_S\sum_{n=1}^{N-1}[R_n]
+2k_I\sum_{n=1}^{N}[L_n]
+k_S\sum_{n=2}^{N-1}[R_nL]
+k_I\sum_{n=2}^{N}[L_nR], 
\label{lambdaR} 
\end{equation}
and $p_1^{(RL)}=p_1^{(LR)}=0$.

\end{article}
\end{document}